\documentstyle[12pt]{article}

\def\hybrid{\topmargin -20pt  \oddsidemargin 0pt
      \headheight 0pt   \headsep 0pt
      \textwidth 6.25in 
      \textheight 9.5in 
      \marginparwidth .875in
      \parskip 5pt plus 1pt   \jot = 1.5ex}

\hybrid

\begin{document}
\def\x{\times}
\def\beq{\begin{equation}}
\def\eeq{\end{equation}}
\def\beqa{\begin{eqnarray}}
\def\eeqa{\end{eqnarray}}
\def\L{ {\cal L}}
\def\C{ {\cal C}}
\def\N{ {\cal N}}
\def\calE{{\cal E}}
\def\lin{{\rm lin}}
\def\Tr{{\rm Tr}}
\def\cF{{\cal F}}
\def\cD{{\cal D}}
\def\modS{{S+\bar S}}
\def\mods{{s+\bar s}}
\newcommand{\Fg}[1]{{F}^{({#1})}}
\newcommand{\cFg}[1]{{\cal F}^{({#1})}}
\newcommand{\cFgc}[1]{{\cal F}^{({#1})\,{\rm cov}}}
\newcommand{\Fgc}[1]{{F}^{({#1})\,{\rm cov}}}
\def\mpl{m_{\rm Planck}}
\def\mxth{\mathsurround=0pt }
\def\xversim#1#2{\lower2.pt\vbox{\baselineskip0pt \lineskip-.5pt
x  \ialign{$\mxth#1\hfil##\hfil$\crcr#2\crcr\sim\crcr}}}
\def\simgr{\mathrel{\mathpalette\xversim >}}
\def\simle{\mathrel{\mathpalette\xversim <}}

\newcommand{\ms}[1]{\mbox{\scriptsize #1}}
\renewcommand{\a}{\alpha}
\renewcommand{\b}{\beta}
\renewcommand{\c}{\gamma}
\renewcommand{\d}{\delta}
\newcommand{\th}{\theta}
\newcommand{\TH}{\Theta}
\newcommand{\pa}{\partial}
\newcommand{\g}{\gamma}
\newcommand{\G}{\Gamma}
\newcommand{\A}{\Alpha}
\newcommand{\B}{\Beta}
\newcommand{\D}{\Delta}
\newcommand{\e}{\epsilon}
\newcommand{\E}{\Epsilon}
\newcommand{\z}{\zeta}
\newcommand{\Z}{\Zeta}
\newcommand{\k}{\kappa}
\newcommand{\K}{\Kappa}
\renewcommand{\l}{\lambda}
\renewcommand{\L}{\Lambda}
\newcommand{\m}{\mu}
\newcommand{\M}{\Mu}
\newcommand{\n}{\nu}
\newcommand{\X}{\Chi}
\newcommand{\R}{\Rho}
\newcommand{\s}{\sigma}
\renewcommand{\S}{\Sigma}
\renewcommand{\t}{\tau}
\newcommand{\T}{\Tau}
\newcommand{\y}{\upsilon}
\newcommand{\Y}{\upsilon}
\renewcommand{\o}{\omega}
\newcommand{\q}{\theta}
\newcommand{\h}{\eta}

\def\dota{ {\dot{\alpha}} }
\def\lag{Lagrangian}
\def\Kahler{K\"{a}hler}
\def\kahler{K\"{a}hler}
\def\A{ {\cal A}}
\def\C{ {\cal C}}
\def\F{{\cal F}}
\def\cL{ {\cal L}}

\def\R{ {\cal R}}
\def\x{ \times }
\def\beq{\begin{equation}}
\def\eeq{\end{equation}}
\def\beqa{\begin{eqnarray}}
\def\eeqa{\end{eqnarray}}

\sloppy
\newcommand{\be}{\begin{equation}}
\newcommand{\eq}{\end{equation}}
\newcommand{\ov}{\overline}
\newcommand{\un}{\underline}
\newcommand{\p}{\partial}
\newcommand{\la}{\langle}
\newcommand{\ra}{\rangle}
\newcommand{\bl}{\boldmath}
\newcommand{\ds}{\displaystyle}
\newcommand{\nl}{\newline}
\newcommand{\Nzahl}{{\bf N}  }
\newcommand{\zzahl}{ {\bf Z} }
\newcommand{\Zzahl}{ {\bf Z} }
\newcommand{\Qzahl}{ {\bf Q}  }
\newcommand{\Rzahl}{ {\bf R} }
\newcommand{\Czahl}{ {\bf C} }
\newcommand{\wt}{\widetilde}
\newcommand{\wh}{\widehat}
\newcommand{\fs}[1]{\mbox{\scriptsize \bf #1}}
\newcommand{\ft}[1]{\mbox{\tiny \bf #1}}
\newtheorem{satz}{Satz}[section]
\newenvironment{Satz}{\begin{satz} \sf}{\end{satz}}
\newtheorem{definition}{Definition}[section]
\newenvironment{Definition}{\begin{definition} \rm}{\end{definition}}
\newtheorem{bem}{Bemerkung}
\newenvironment{Bem}{\begin{bem} \rm}{\end{bem}}
\newtheorem{bsp}{Beispiel}
\newenvironment{Bsp}{\begin{bsp} \rm}{\end{bsp}}
\renewcommand{\arraystretch}{1.5}

\addtocounter{section}{1}

\renewcommand{\thesection}{\arabic{section}}
\renewcommand{\theequation}{\thesection.\arabic{equation}}

\parindent0em

\def\S4{\frac{SO(4,2)}{SO(4) \otimes SO(2)}}
\def\P3{\frac{SO(3,2)}{SO(3) \otimes SO(2)}}
\def\MGd{\frac{SO(r,p)}{SO(r) \otimes SO(p)}}
\def\SOd{\frac{SO(r,2)}{SO(r) \otimes SO(2)}}
\def\SO2{\frac{SO(2,2)}{SO(2) \otimes SO(2)}}
\def\SUm{\frac{SU(n,m)}{SU(n) \otimes SU(m) \otimes U(1)}}
\def\SUS{\frac{SU(n,1)}{SU(n) \otimes U(1)}}
\def\SK{\frac{SU(2,1)}{SU(2) \otimes U(1)}}
\def\SU{\frac{ SU(1,1)}{U(1)}}


\begin{titlepage}
\begin{center}
\hfill CERN-TH/96-232\\
\hfill HUB-EP-96/49\\
\hfill {\tt hep-th/9609111}\\

\vskip .3in

{\bf ON THE DUALITY BETWEEN THE HETEROTIC STRING\\ AND F-THEORY
IN 8 DIMENSIONS}

\vskip .2in

{\bf Gabriel Lopes Cardoso$^a$, Gottfried Curio$^b$, \\ 
Dieter L\"ust$^b$ and Thomas Mohaupt$^b$}\footnote{ 
Email: \tt cardoso@surya11.cern.ch, curio@qft2.physik.hu-berlin.de, 
\hfill \\
luest@qft1.physik.hu-berlin.de, 
mohaupt@qft2.physik.hu-berlin.de}
\\
\vskip 1.2cm

$^a${\em Theory Division, CERN, CH-1211 Geneva 23, Switzerland}\\
$^b${\em Humboldt-Universit\"at zu Berlin,
Institut f\"ur Physik, 
D-10115 Berlin, Germany}\\

\vskip .1in

\end{center}

\vskip .2in

\begin{center} {\bf ABSTRACT } \end{center}
\begin{quotation}\noindent
In this note we compare the moduli spaces of the heterotic
string compactified on a two--torus and F--Theory compactified on
an elliptic K3 surface for the case of an unbroken
$E_8 \times E_8$ gauge group. 
The explicit map relating the deformation parameters 
$\alpha$ and $\beta$ of the F--Theory K3 surface 
to the moduli $T$ and $U$ of the heterotic torus is found
using the close relationship between the K3 discriminant
and the discriminant of the Calabi--Yau--threefold $X_{1,1,2,8,12}(24)$
in the limit of a large base ${\bf P}^1$.

\end{quotation}
\vskip 4cm
September 1996\\
\hfill CERN-TH/96-232 \\
\end{titlepage}
\vfill
\eject

\newpage

During the last year it has become likely
that a non--perturbative
formulation of string theory (or of the theory behind it) requires the
introduction of additional dimensions beyond the critical dimension
10 of perturbative supersymmetric string theories. From the space--time
point of view all such constructions have in
common that they enable one to regard the dilaton as a geometric
modulus arizing from compactifying these extra dimensions.

There are at time two constructions of this type called 
M--theory \cite{MRev}
and F--theory \cite{Vafa} - \cite{Sen}.
More specifically, in \cite{Vafa} Vafa conjectured
 that II B superstring theory should
be regarded as the toroidal compactification of twelve--dimensional
F--theory. Adapting the stringy cosmic string construction 
\cite{CosmStr} new compactifications of the II B strings
on D--manifolds were constructed 
in which the complexified coupling varies over the
internal space. These compactifications then have a beautiful
geometric interpretation as compactifications of F--theory on
elliptically fibred manifolds, where the fibre encodes the behaviour of
the coupling, the base is the D--manifold, and the points where
the fibre degenerates specifies the positions of the D--branes 
in it. Moreover compactifications of F--theory on elliptic Calabi--Yau
twofolds (the K3), threefolds and fourfolds can be argued to be
dual to certain heterotic string theories in 8,6 and 4 dimensions and
have provided new insights into the relation between geometric 
singularities and perturbative as well as non--perturbative 
gauge symmetry enhancement and into the structure of moduli spaces.

The simplest but already instructive example of a 
F--theory compactification is given by the compactification
to 8 dimensions on a K3 which is believed to be dual to the
heterotic string on a two--torus. It is the purpose of this note
to make explicit the precise relation between a specific part of
the moduli spaces of both theories.

In the next section we will
first  consider the K3 discriminant with special
emphasis on its singular points which are precisely related to the
points of enhanced gauge symetries for particular values of the
heterotic $T_2$ moduli.
In the subsequent section we will work out the precise relation between
the heterotic $T_2$ moduli $T$, $U$ and the corresponding $F$-theory
$K3$ moduli $\alpha$, $\beta$. 
For this purpose we will use the four-dimensional 
duality \cite{KV} - \cite{CCLM} between the heterotic
string on $T_2^{T,U}\times K3^H$ with special $SU(2)$ 
instanton embedding 
and the type IIA string on a Calabi--Yau-threefold $CY$. The relevant
$CY$  
is a $K3$ fibration, i.e. locally we can write $CY={\bf P}_1\times K3^{CY}$.
The four-dimensional duality between the heterotic string and F--theory
on  $CY\times T_2^F$ is then implied 
by following the arguments given in \cite{Vafa,MV1}, where  
the considered $CY$ is at the same time
an elliptic fibration.
For concreteness we choose $CY$ to be given by $X_{1,1,2,8,12}(24)$,
with Hodge numbers $h_{1,1}=3$ and $h_{2,1}=243$.
This leads to three vector moduli $S$, $T$ and $U$, 
where all other gauge symmetries are broken by generic values of the
hyper multiplets.

Now we consider the decompactification limit to six dimensions
by making the  base ${\bf P}_1$ large, which is common to all
three models (regarding $K3^H$ locally as ${\bf P}_1\times T_2^{K3}$).
In this way,
using the (reverse) adiabatic argument of \cite{VaWi},
 we are  dealing with a heterotic string
on $T_2^{T,U}\times T_2^{K3}$ where the non-Abelian gauge
symmetries $E_8\times E_8$ 
are now broken by the Wilson line vector multiplets.
The six-dimensional heterotic string is in turn dual to the type IIA string on
$K3^{CY}$ or respectively dual to
F--theory on $T_2^F\times K3^{CY}$. 
Now it is important to remember that, in the limit of large ${\bf P}_1$,
 the heterotic moduli $T$ and $U$ can by
explicitly related to the $K3^{CY}$ moduli via the mirror map
from IIA to IIB. Finally,
we can trade $T_2^F$ for $T_2^{K3}$, and we can
directly compare the 8-dimensional heterotic string on $T_2^{T,U}$
(with non-vanishing Wilson lines and completely broken $E_8\times E_8$
gauge symmetries)
to F--theory on $K3^{CY}$.  
In this way we will  obtain the 
exact relation between
the heterotic  moduli $T$, $U$ and the $K3^{CY}$ moduli. 
Note that so far F--theory on $K3^{CY}$ has broken $E_8\times E_8$
gauge symmetries. We will explain  how we relate $K3^{CY}$ to
a different $K3$ with $E_8\times E_8$ singularities and hence
unbroken $E_8\times E_8$ gauge symmetries.

\section*{The K3 discriminant and its relation to heterotic moduli}

Vafa has argued \cite{Vafa}
that the heterotic string compactified
on a two--torus in the presence of Wilson lines 
is dual to F--theory compactified on the family 
\be
y^2 = x^3 + f^{(8)}(z) x + f^{(12)}(z)
\label{K3withWL}
\eq
of elliptic K3 surfaces, where $f^{(k)}(z)$ is a polynomial
of order $k=8,12$ respectively.
In particular F--theory on the two parameter subfamily
\be
y^2 = x^3 + \alpha z^4 x + (z^5 + \beta z^6 + z^7)
\label{K3withoutWL}
\eq
of K3's with $E_8$ singularities at $z=0,\infty$
is dual to the heterotic theory with Wilson lines switched off \cite{MV2}.
Therefore there must exist a map which relates the complex structure and
K\"ahler moduli $U$ and $T$ of the torus on which the heterotic theory
is compactified to the two
complex structure moduli $\alpha$ and $\beta$ in
(\ref{K3withoutWL}). We will work out this map explicitly.

Let us first recall that it was claimed in \cite{MV2} 
that in a certain
limit the K3 fibre becomes constant over the base and moreover
is then identical to the heterotic torus. Here ``being identical''
means having the same complex structure, since the K\"ahler modulus
of the elliptic fibre of the K3 is frozen. 

On the heterotic side the limit to be considered is 
the decompactification limit
in which the K\"ahler modulus $T$ is sent to infinity.
On the F--theory side one sends both $\alpha$ and $\beta$ to infinity,
keeping the ratio $\frac{\alpha^3}{\beta^2}$ fixed. This has the effect
that the complex structure of the fibre becomes constant away
from $z=0,\infty$. 
To make this explicit note that the complex structure modulus $\tau_z$ 
of the fibre can be 
read off from the cubic equation (\ref{K3withWL}) to be
\be
j(\tau_z) = 
\frac{ 1728 }{ 1 + 
\frac{27}{4} \frac{ ( f^{(12)}(z) )^2}{ (f^{(8)} (z))^3 } }.
\eq
Using the special form (\ref{K3withoutWL}) of the coefficents
$f^{(k)}(z)$ and identifying the limit 
$\alpha, \beta \rightarrow \infty$, $\frac{\alpha^3}{\beta^2}$ finite
with $T \rightarrow \infty$, $U$ arbitrary, the prediction is that 
\be
j(iU) = \lim_{\alpha, \beta \rightarrow \infty} j(\tau_z) =
\frac{ 1728 }{ 1 + 
\frac{27}{4}
\frac{ \beta^2}{ \alpha^3 } },
\;\;\;\mbox{if}\;\;\;
j(iT) = \infty.
\label{TauAlphaBeta}
\eq
We will show later on that this is indeed the case.

In order to get more information about the relation of
$(T, U)$ to $(\alpha, \beta)$ we now make use of the fact
that for the heterotic string on a two--torus the generic gauge group
$U(1)^2$ is enhanced to $SU(2) \times U(1)$, $SU(2)^2$ and
$SU(3)$ for $T=U$, $T = U = 1$ and 
$T = U = e^{2 \pi i/12}$ 
(neglecting the $E_8^2$ which is already present in
ten dimensions and unbroken due to the absense of Wilson lines). 
In F--theory these gauge symmetry enhancements must
arise from singularities of the K3 surface at special values
of the parameters $\alpha, \beta$. In our example these
singularities must be of type $A_1$, $A_1^2$ and $A_2$ 
(neglecting the two $E_8$ singularities  which are present for
all values of $\alpha$ and $\beta$ at $z=0,\infty$). 

Our next step is to compute at which values of $\alpha$ and
$\beta$ these singularities occur on the K3. To do so we start
from the defining equation of the surface
\be
F(x,y,z) = x^3 - y^2 + \alpha \, z^4 \,x +  z^5 + \beta z^6 + z^7= 0
\label{DefEqu}
\eq
and look for singularities by solving $F_x = F_y = F_z =0$. 
Substituting these potential singular points back into the
defining equation (\ref{DefEqu}) gives a relation between the
parameters which takes the form $\Delta(\alpha, \beta) =0$,
where $\Delta^{(K3)} = \Delta(\alpha, \beta)$ is the discriminant of the
surface. In our case one finds
\be
\Delta^{(K3)} = \left( \alpha^3 + \frac{27}{4} \beta^2 + 27 \right)^2 
- 27^2 \beta^2,
\label{K3discriminant}
\eq
which can be factorized as
$\Delta^{(K3)} = \prod_{i} (\beta - \beta_i)$, where
$\beta_{\pm 1,\pm 1} =  \pm 2 
\pm \frac {2\,\sqrt{3}}{9} (-\alpha)^{3/2}$.
As long as both $\alpha$ and $\beta$ are not zero the four 
zeros of the discriminant are distinct. Along each of the
four branches $\beta = \beta_{\pm 1, \pm 1}(\alpha)$ there is
precisely one singular point $(x_0, y_0, z_0)$ (outside $z=0, \infty$)
which is located at $\left( \epsilon_2 \sqrt{-\frac{\alpha}{3}},\;
0,\;
- \epsilon_1 \right)$ for $\beta = \beta_{\epsilon_1, \epsilon_2}$
(,where $\epsilon_{1,2} = \pm 1$).

Since the matrix of second derivatives is non--degenerate for
$\alpha \not= 0$ ($\det(F_{ij}) \sim \sqrt{-\alpha}$)
the singularity is of type $A_1$, because one can redefine coordinates such
that the surface is locally given by
$F(x,y,z) = (x-x_0)^2 + (y-y_0)^2 + (z-z_0)^2 =0$
near the critical point, which is the standard form of an $A_1$ 
singularity.

Let us then discuss what singularities appear if two zeros of
the discriminant coincide, starting with the case $\beta=0$.
This implies $\alpha^3 = -27$. Taking for example the solution
$\alpha = -3$ we find that it corresponds to an intersection
of the branches $\beta_{+1,-1}$  and $\beta_{-1,+1}$, which shows
that there are simultanously two critical points located at
$(-1, \;0, \;-1)$ and 
$(1, \; 0, \; 1)$. 
Both are of type $A_1$ because 
$\det(F_{ij}) \sim \sqrt{-\alpha} \not= 0$.
The same happens for the other two solutions
$\alpha = 3( \frac{1}{2} \pm \frac{i}{2} \sqrt{3}) $
of $\alpha^3 = - 27$ with different localizations of the critical
points.

The other case is $\alpha=0$ which implies $\beta^2 =4$.
The two subcases $\beta = \pm 2$ correspond to intersections
of branches $\beta_{+1,+1} = \beta_{+1,-1}$ and $\beta_{-1,+1} = 
\beta_{-1,-1}$.
This time there is only one singular point located at
$(0, \; 0, \; -1)$ for $\beta=2$ and
$(0,\; 0,\; 1)$ for $\beta=-2$.
The singularity is not of type $A_1$, because $\alpha=0 \Rightarrow
\det ( F_{ij} ) = 0$. At the critical point $(x_0,y_0,z_0)$ one computes 
$F_{xx} = 0$,
$F_{xxx} \not=0$,
$F_{yy} \not= 0$,
$F_{zz} \not= 0$, and
$F_{ij} = 0$, if $i\not=j$.
Therefore 
the surface can be locally brought to the form
$F(x,y,z) =  (x - x_0)^3 + (y - y_0)^2 + (z - z_0)^2 = 0$
making explicit that the singularity is of type $A_2$.

The appearence of surface singularities is related to the location
of singular fibres as is well known from the stringy cosmic
string construction \cite{CosmStr}. In general the locations of degenerate
fibres are obtained by solving
\be
j(\tau_z) = 1728 \cdot
\frac{4 \cdot (f^{(8)}(z))^3  }{ 4 \cdot (f^{(8)}(z))^3 + 
27 \cdot (f^{(12)}(z))^2  } = \infty 
\label{SingularFibre}
\eq
for $z$. For generic moduli this is equivalent to solving
\be
\Delta^{(T)} =4 \cdot (f^{(8)}(z))^3 + 27 \cdot (f^{(12)}(z) )^2   = 0,
\eq
where $\Delta^{(T)}$ is the discriminant of the elliptic fibre.
In general this equation has 24 distinct solutions
corresponding to 24 non--coinciding singular fibres.
Restricting to the two--parameter family (\ref{K3withoutWL})
and ignoring singular fibres over $z=0, \infty$ this
becomes
\be
4 \alpha^3 z^2 + 27 ( z^2 + \beta z + 1)^2 = 0.
\eq
For generic $\alpha, \beta$ there are four distinct roots
corresponding to four singular fibres over four different points
on the base. On the discriminant locus $\Delta(\alpha,\beta)=0$
two of the four roots coincide , 
whereas at the $A_1^2$ point
the four roots combine into two pairs. Finally at
the $A_2$ point all four roots coincide. This gives a nice 
explicit example of how singularities come about in the stringy
cosmic string construction of K3: whereas isolated singular fibres
give regular points, 
all kinds of $ADE$ singularities can be obtained by letting singular
fibres coincide in a particular way \cite{CosmStr}. 

The discriminant $\Delta^{(K3)}$ (\ref{K3discriminant})
of the K3 surface and the discriminant $\Delta^{(T)}$
of its elliptic fibre are closely related. One can check that $\Delta^{(K3)}$
is the discriminant in the usual (algebraic) sense
of the discriminant $\Delta^{(T)}$ of the elliptic fibre: 
$\Delta^{(K3)}$ vanishes if two or more
zeros of $\Delta^{(T)}$ coincide, reflecting the fact that 
K3 singularities do not come from singular fibres but from coinciding
singular fibres. This is analogous to the relation between
K3 fibred Calabi--Yau threefolds and their K3 fibres which were
discussed in \cite{KLM}, \cite{KLMVW}.  
Moreover there is a simple asymptotic relationship
between the two discriminants in the limit $\alpha, \beta \rightarrow 
\infty$, $\frac{\alpha^3}{\beta^2}$ finite:
\be
\frac{\Delta^{(K3)}}{\alpha^6} = \left( 
\frac{\Delta^{(T)}}{\alpha^3}
\right)^2 + O \left( \frac{\beta^2}{\alpha^6}, \frac{1}{\alpha^3}  \right) .
\eq
Again a similar relation was observed in the Calabi--Yau context
in \cite{KLM}, \cite{KLMVW}.

Summarizing we have found the critical values for $\alpha$, $\beta$
where the K3 surfaces (\ref{K3withoutWL}) devellop $A_1$, $A_1^2$
or $A_2$ singularities. These values must be mapped to the corresponding
critical values of the heterotic moduli $T$, $U$:
\be
\begin{array}{lclcl}
T = U &\leftrightarrow& j(iT) = j(iU) 
&\leftrightarrow& \Delta^{(K3)} =\Delta(\alpha, \beta) = 0\\
T = U =1 &\leftrightarrow& j(iT) = j(iU) =1728
&\leftrightarrow& \alpha^3 = -27,\; \beta=0\\
T = U =e^{2 \pi i /12} &\leftrightarrow& j(iT) = j(iU) =0
&\leftrightarrow& \alpha = 0,\; \beta^2 = 4.\\
\label{TableCritVal}
\end{array}
\eq
We also expect to find the asymptotic relation (\ref{TauAlphaBeta}) 
in the limit $j(iT) \rightarrow \infty$,
$j(iU)$ finite corresponding to $\alpha, \beta \rightarrow \infty$,
$\frac{\alpha^3}{\beta^2}$ finite. Finally note that on the heterotic
side mirror symmetry exchanges $T$ and $U$. Therefore one might 
expect that $\alpha$ and $\beta$ are given by symmetric combinations
$j(iT)\cdot j(iU)$ and $j(iT) + j(iU)$ as  the case of the map
between the $(T,U)$ moduli space and the weak coupling limit
of the conifold locus of the Calabi--Yau threefold $X_{1,1,2,8,12}(24)$ 
\cite{KV}, \cite{KLM}. In fact we can use that map in order
to relate $(T,U)$ to $(\alpha, \beta)$, as we will explain in
the next section.

\section*{Relation to the $S$-$T$-$U$ Calabi Yau}
As argued in the introduction,
one should be able to relate 
$\alpha$ and $\beta$ to
$T$ and $U$, by using some of the results
of \cite{KLM} for the $S$-$T$-$U$ Calabi--Yau $X_{1,1,2,8,12}(24)$.
A priori, one might have thought that this isn't possible, because
$F$--theory on K3 exhibits $E_8$--type singularities, whereas the 
$S$-$T$-$U$ Calabi--Yau doesn't.  The point, however, is that
in $F$--theory there is a clear distinction between moduli of the 
$K3$, and moduli of the $E_8 \times E_8$.
The perturbative gauge symmetry enhancement on the heterotic
side along $T=U$ is controlled by the ``middle polynomials"
$xz^4$ and $z^6$
on the
$F$--theory side, whereas the ``lower and higher polynomials"
are related to enhancement to $E_8 \times E_8$ or a subgroup thereof.
Thus, the idea is to first obtain the "middle polynomials" for the 
$S$-$T$-$U$ Calabi--Yau $X_{1,1,2,8,12}(24)$, and then to identify them with
the "middle polynomials" of $F$--theory compactified on $K3$.
By doing so, one obtains an explicit map between 
the $F$--theory moduli $\alpha$ and $\beta$ and
the heterotic moduli $T$ and $U$.  Some of the discussion given on page 5
of \cite{BIKMSV} seems to be pointing into the same direction.
The defining polynomial for the Calabi--Yau $X_{1,1,2,8,12}(24)$
is, according to eq. (7) of \cite{KLM}, given by
\beqa
p=x_1^{24} + x_2^{24} + x_3^{12} + x_4^3 + x_5^2 - 12 \psi_0 x_1x_2x_3x_4x_5
- 2 \psi_1(x_1x_2x_3)^6 - \psi_2(x_1x_2)^{12}.
\eeqa

In order to show that this Calabi--Yau is a 
$K3$--fibration, we set $x_2 = \lambda x_1$, $\tilde{x}_1 = x_1^2$ and obtain
\beqa
p_{K3}=x_3^{12} + x_4^3 + x_5^2 + (1+\lambda^{24} - \psi_2 \lambda^{12}) 
\tilde{x_1}^{12} - 
12 \psi_0 \lambda \tilde{x}_1x_3x_4x_5
- 2 \psi_1 \lambda^6 (\tilde{x}_1x_3)^6.
\label{k3fiber}
\eeqa
This describes the $K3$ fiber of the Calabi--Yau
$X_{1,1,2,8,12}(24)$.  We would like to write it in the form 
of a Weierstrass equation, that is in the form $y^2=x^3+xf^{(8)}(z) 
+ f^{(12)}(z)$.
With $x_5=y, x_4=x,  x_3=z , \tilde{x}_1 = w$ and renaming 
$x \rightarrow - x$ the equation $p_{K3}=0$ turns into
\beqa
y^2 + ( 12 \psi_0 \lambda z w ) yx = x^3 - z^{12}
-(1+\lambda^{24}-\psi_2 \lambda^{12}) w^{12} + 2 \psi_1 \lambda_6 (z w)^6.
\label{ynw}
\eeqa
This is precisely of the form of eq. (3.2) in \cite{BIKMSV},
with the following identifications
\beqa
a_1 &=& 12 \psi_0 \lambda z w \;\;,\;\;
a_2=a_3=a_4=0 \nonumber\\
a_6 &=& -z^{12}-(1+\lambda^{24}-\psi_2 \lambda^{12})w^{12}
+ 2 \psi_1 \lambda^6 (zw)^6.
\eeqa
The coefficients $b_j$ of eq. (3.3) of \cite{BIKMSV} 
are then given by $b_2 = a_1^2 , b_4 = 0, b_6 = 4 a_6,b_8 = b_2 a_6$.
The Weierstrass form can now be obtained 
by completing the square in $y$ and then completing the cube in $x$.
The resulting functions $f^{(8)}$ and $f^{(12)}$ are then given as 
follows (eq. (3.4)
of \cite{BIKMSV})
\beqa
f^{(8)}&=&-\frac{1}{48}b_2^2=-\frac{1}{48}a_1^4=\alpha_{CY}
 z^4 w^4 \nonumber\\
f^{(12)} &=& \frac{1}{864} (b_2^3 + \frac{1}{216} b_6) = \frac{1}{864}
(a_1^6 +  864 a_6)\nonumber\\
&=& - z^{12} -\gamma w^{12}
+\beta_{CY} z^6 w^6.\label{fcy}
\eeqa

with $\alpha_{CY}=-432(\psi_0\lambda)^4,\beta_{CY}=3456(\psi_0 \lambda)^6+
2\psi_1\lambda^6$ and $\gamma=1 + \lambda^{24} - \psi_2 \lambda^{12}$.
Note that $f^{(8)}$ and $f^{(12)}$ have degrees $8$ and $12$, 
respectively, in $z$ and $w$ 
(in the following we will work in the chart $w=1$).
Note that $f^{(8)}(z)$ contains precisely (and only) a $z^4$-term, whereas
$f^{(12)}(z)$ contains a $z^6$-term, but no $z^5$ or $z^7$ term and, hence,
no singularities of the $E_8$ type, as should be the case
for the Calabi--Yau $X_{1,1,2,8,12}(24)$.  Instead, $f^{(12)}(z)$ contains
a constant $z^{12}$-term as well as a $z^0$-term.  
Next, consider rewriting $f^{(8)}(z)$ and $f^{(12)}(z)$ in terms of the
complex structure moduli $\bar{x}$ and $\bar{z}$ (not to be confused
with the earlier coordinates $x,z$),
given in \cite{KLM}: $\bar{x} =-\frac{\psi_1}{3456 \psi_0^6} \;,
\bar{z} =-\frac{\psi_2}{\psi_1^2}$.

Thus

\beqa
\alpha_{CY} ^3=\frac{27}{4}\psi_2\lambda^{12}\frac{1}{\bar{x}^2\bar{z}},\qquad
\beta_{CY} ^2=-\psi_2\lambda^{12}\frac{(1-2\bar{x})^2}{\bar{x}^2\bar{z}}.
\eeqa

Now we consider the weak coupling limit $\psi_2\rightarrow\infty$,
i.e. the limit of large ${\bf P}_1$.
With 
\beqa
\bar{x}&=&
864 \frac{j(iT) + j(iU) - 1728}
{j(iT) j(iU) +\sqrt{j(iT) (j(iT) - 1728 )}\sqrt{j(iU) (j(iU) - 1728)}},
\nonumber\\
\bar{z}&=&864^2\frac{1}{j(iT) j(iU) \bar{x}^2}\label{zxTU},
\eeqa
one finds, using the relation $4\bar{x}(1-\bar{x})=
1728\frac{j(iT) + j(iU) - 1728}{j(iT) j(iU)}$,
 for
the combination
\beqa
\frac{\beta_{CY}^2}{\alpha_{CY}^3} = - \frac{4}{27} (1-2\bar{x})^2
=- \frac{4}{27}(1-1728\frac{j(iT)+j(iU)-1728}{j(iT)j(iU)}).
\eeqa

So finally one has 
\beqa
j(iT) j(iU) &=&\;\;\frac{1728^2}{\psi_2\lambda^{12}}
\frac{\alpha_{CY}^3}{27},\nonumber\\
(j(iT) - 1728)(j(iU) - 1728)&=&-\frac{1728^2}{\psi_2\lambda^{12}}
\frac{\beta_{CY}^2}{4}.
\label{Result}
\eeqa

Let us now interpret the above result in the light of our original 
question. For this we have to go from the K3 in eq.(\ref{K3withoutWL})
(in homogenized form),

\be
y^2=x^3+\alpha z^4w^4x+z^5w^7+\beta z^6w^6+z^7w^5,
\eq

to the K3 of the $S$-$T$-$U$ Calabi--Yau, defined
in eq.(\ref{fcy}),

\be
y^2=x^3+\alpha_{CY} z^4w^4x-\gamma w^{12}+\beta_{CY} z^6w^6-z^{12},
\eq

by the redefinition 
$z\rightarrow \rho (z,w)z$, $w\rightarrow \sigma \frac{1}{\rho (z,w)}w$.
This leads to the condition $\sigma^{12}=\gamma$ and

\be
\alpha^3_{CY}=\gamma \alpha^3,\;\; \beta^2_{CY}=\gamma \beta^2.
\eq 
Replacing $\alpha_{CY}$, $\beta_{CY}$ by $\alpha$, $\beta$ corresponds
to the transition from non-vanishing 
heterotic Wilson lines with broken $E_8\times
E_8$ to the case of vanishing Wilson lines with unbroken $E_8\times E_8$,
which was the starting point in eq.(\ref{K3withoutWL}).
So the discriminant for the locus $T=U$ we got from the 
$T,U \leftrightarrow \alpha_{CY},\beta_{CY}$
matching above is given by

\beqa
(j(iT)-j(iU))^2
&\sim&((-\frac{1}{\psi_2\lambda^{12}})\frac{\alpha_{CY}^3}{27}+
(-\frac{1}{\psi_2\lambda^{12}})\frac{\beta_{CY}^2}{4}-1)^2+
4(-\frac{1}{\psi_2\lambda^{12}})\frac{\alpha_{CY}^3}{27}\nonumber\\
&=&((-\frac{1}{\psi_2\lambda^{12}})\frac{\alpha_{CY}^3}{27}+
(-\frac{1}{\psi_2\lambda^{12}})\frac{\beta_{CY}^2}{4}+1)^2-
4(-\frac{1}{\psi_2\lambda^{12}})\frac{\beta_{CY}^2}{4}\nonumber\\
&=&((-\frac{\gamma}{\psi_2\lambda^{12}})\frac{\alpha^3}{27}+
(-\frac{\gamma}{\psi_2\lambda^{12}})\frac{\beta^2}{4}+1)^2-
4(-\frac{\gamma}{\psi_2\lambda^{12}})\frac{\beta^2}{4}.
\eeqa

This is precisely proportional to the 
discriminant eq.(\ref{K3discriminant}) found in the previous section
as in the weak 
coupling limit $\psi_2\rightarrow\infty$ one has $\gamma
\rightarrow -\psi_2\lambda^{12}$.

In summary, let us display our main result which explicitly
relates the moduli $T$ and $U$ of the heterotic string compactified
on $T_2$ to the moduli $\alpha$ and $\beta$ of $F$-theory on $K3$:
\beqa
j(iT) j(iU) &=&-1728^2\frac{\alpha^3}{27},\nonumber\\
(j(iT) - 1728)(j(iU) - 1728)&=&\;\; 1728^2\frac{\beta^2}{4}.
\label{Resulta}
\eeqa

As as useful check of this results consider the ratio ${\beta^2\over \alpha^3}$
in the limit $T\rightarrow\infty$:
\beqa
{\beta^2\over \alpha^3}\rightarrow -{4\over 27}(1-{1728\over j(iU)}).
\eeqa
This is in precise agreement with eq.(\ref{TauAlphaBeta}), and we therefore
 find that in the limit $\alpha,\beta\rightarrow \infty$ 
and $\frac{\beta^2}{\alpha^3}$ fixed,
the fibre modulus  equals
the heterotic modulus $iU$
in accordance with \cite{MV2}. 

Let us finally remark that our result can be regarded as the two-parameter
generalization of the one-parameter torus
$y^2 = x^3 + \alpha x + \beta$. 
Comparing this with the
well-known
Weierstrass form ($g_2=\frac{4}{3}\pi^4 E_4$, $g_3=
\frac{8}{27}\pi^6 E_6$)
\be
y^2 = 4 x^3 - g_2(\tau) x - g_3(\tau)
\eq
yields $\alpha = -\frac{1}{3}\pi^4E_4$, $\beta =-\frac{2}{27}\pi^6E_6$;
so in this case 
one has with $\Delta=\frac{(2\pi)^{12}}{1728}(E_4^3-E_6^2)$ that
\beqa
j(\tau)\Delta(\tau)&=&-1728^2\frac{\alpha^3}{27},\nonumber\\
(j(\tau)-1728)\Delta(\tau)&=&1728^2\frac{\beta^2}{4}.\label{jtaual}
\eeqa

\end{document}